# Compact All-optical Reservoir Computing *via* Luminescence Dynamics in Rare-earth Ions-doped Nanocrystals

Junyan Chen[1,†], Jingsong Fu[1,†], Jie Xu[1], Yixiang Qin[1], Axin Du[1], Kaiyang Wang[1,*], Limin Jin[1,2,*], Can Huang[1,3,4,*]

[1] Ministry of Industry and Information Technology Key Lab of Micro-Nano Optoelectronic Information System, Guangdong Provincial Key Laboratory of Semiconductor Optoelectronic Materials and Intelligent Photonic Systems, Harbin Institute of Technology, Shenzhen 518055, China.

[2] National Key Laboratory of Science and Technology on Advanced Composites in Special Environments, Harbin Institute of Technology, Harbin 150080, China.

[3] Quantum Science Center of Guangdong-Hongkong Macao Greater Bay Area, Shenzhen 518055, China.

[4] Heilongjiang Provincial Key Laboratory of Advanced Quantum Functional Materials and Sensor devices, Harbin Institute of Technology, Harbin 150001, China

[†] These authors contribute equally to this work.

*Corresponding author: Can Huang: huangcan@hit.edu.cn; Limin Jin: jinlimin@hit.edu.cn; Kaiyang Wang: optoelectrogump@163.com

**Abstract**

Optical neuromorphic computing offers a promising route to high-speed, energy-efficient information processing. However, photonic neurons, as the critical components for enhancing computational expressivity, still face significant bottlenecks in nonlinear mapping and memory capacity. Here, we demonstrate an all-optical reservoir computing system based on rare-earth ions-doped nanocrystals for the first time, leveraging their intrinsic nonlinear luminescence dynamics and multi-timescale memory. Unlike traditional schemes that require bulky optical delays or intricate resonant structures, our platform exploits the material's inherent properties: nonlinear cross-relaxation processes enable nonlinear mapping while millisecond-scale metastable energy levels provide fading memory. As a proof of concept, we achieve 90.7% accuracy in MNIST digit classification and low-error chaotic time-series prediction (NRMSE < 0.1) using the rare- earth ions based system. Our work significantly reduce system footprint and complexity, offering a scalable, fully optical solution for edge computing and real-time neuromorphic applications.

**Introduction**

The rapid development of the Internet of Things (IoT) and big data applications has imposed increasing demands on sensing and computing systems, necessitating intelligent hardware architectures that combine compact size, low power consumption, and edge computing capabilities. Traditional computing paradigms face significant challenges in terms of energy efficiency and real-time performance, driving the emergence of novel bio-inspired computing approaches such as neuromorphic computing [1-3]. Among these, reservoir computing (RC), as an efficient variant of recurrent neural networks (RNNs), has demonstrated unique advantages in temporal signal processing due to its low training cost and strong adaptability [4-6]. Unlike conventional RNNs, RC only requires training the readout layer weights while keeping the random connections within the reservoir fixed, significantly reducing computational complexity. More importantly, RC can be physically implemented using any system exhibiting nonlinear dynamics and short-term memory properties, opening vast opportunities for exploring novel computing media [7-9].

Optical computing has emerged as a pivotal pathway to overcome the limitations of electronic computing, leveraging its inherent advantages of ultrahigh speed, low latency, and natural parallelism [10-12]. Optical neuromorphic computing, in particular, combines the high bandwidth of optical hardware with the energy-efficient characteristics of biological neural networks, offering a novel approach for constructing high-performance artificial intelligence systems. However, existing optical reservoir computing (RC) schemes often rely on complex fiber-optic architectures [13], microring resonators [14], or spatial scattering structures [15], which suffer from bulky system footprints, poor stability, and weak nonlinearity [16,17], severely limiting their applicability in edge computing scenarios.

To address these challenges, here we propose an all-optical reservoir computing platform constructed using rare-earth ion-doped (($Re^{3+}$) nanocrystals, which we demonstrate for the first time experimentally in neuromorphic computing applications. $Re^{3+}$ ions possess rich multi-level energy structures and complex excited-state dynamics [18, 19]. The interaction between radiative transitions and non-radiative cross-relaxation (CR) processes results in highly nonlinear optical responses. Notably, the metastable energy levels of $Re^{3+}$, with lifetimes extending up to milliseconds, offer intrinsic multi-timescale memory capabilities, eliminating the need for external delay feedback to meet the fading memory requirements of reservoir computing [20, 21]. Based on this, we utilized this material to construct an all-optical reservoir computing system, which exhibited superior performance in both the Modified National Institute of Standards and Technology (MNIST) handwritten digit classification task (achieving approximately 90.7% accuracy) and the Mackey-Glass chaotic time series prediction task (with normalized root mean square error, NRMSE < 0.1). Compared to traditional optical RC implementations [13-15], our approach capitalizes on the inherent nonlinear dynamics of $Re^{3+}$ ions without the need for complex optical feedback architectures or any electrical excitation or additional electro-optic modulators, significantly reducing system complexity. Moreover, these nanomaterials demonstrate compatibility with functional micro/nanophotonic components, thus creating the possibility for monolithic integration. Collectively, these features present new opportunities for the development of next-generation low-power edge computing devices.

**Results**

Figure **1a** illustrates the schematic design of the all-optical reservoir network implementation based on luminescence dynamics of $Re^{3+}$ ions. In $Re^{3+}$-doped luminescent materials, the interplay between radiative transitions and non-radiative CR processes establishes a highly nonlinear relationship between the material's luminescence and incident optical excitation. Meanwhle the prevalent long-lived intermediate energy states provide intrinsic temporal memory capability. These characteristics collectively facilitate the construction of physical reservoir layer. As illustrated in Figure **1c**, when the interval between consecutive optical pulses is shorter than the carrier relaxation time of $Re^{3+}$, the $Re^{3+}$ can store input information from previous nodes as hidden states. These hidden states *h(t)* jointly determine the reservoir output *y(t)* with new input data *x(t)*. The influence of preceding virtual nodes persists in the system and gradually decays over time, thereby establishing interconnections among virtual nodes. It is particularly noteworthy that this fading memory effect is crucial for achieving superior RC performance [22]. Owing to the inherent strong memory effects, the system operates without requiring any feedback loops, thereby significantly reducing system's complexity while eliminating the speed limitations imposed by feedback circuits.

As a proof of concept, we chose $Yb^{3+}/Tm^{3+}$ co-doped multi-shell $NaYF_4$ nanocrystals as a model system, taking advantage of their low phonon energy, high infrared absorption, and superior upconversion intensity. The $NaYF_4:Gd@NaYbF_4:Tm@NaYF_4$ core-shell-shell upconversion nanocrystals (UCNCs) were synthesized via a modified coprecipitation method [23,24]. As shown in Figure **1a**, the synthetic process involved: (1) $NaYF_4$:Gd(10 mol%) core nanocrystals (~14 nm) serving as nucleation seeds to guide the growth of $NaYbF_4$:Tm(5 mol%) gain layer (~5 nm). Note that, the $Gd^{3+}$(10 mol%) ions was introduced for the small and uniform core nanocrystals. And (2) an outermost $NaYF_4$ passivation shell (~6 nm) that effectively minimizes detrimental energy dissipation at the surface defects and ligand of nanoparticles, thereby reducing non-radiative quenching [24,25].

This sandwiched multi-shell architecture integrates two critical advances: First, the spatial confinement effect enables $Yb^{3+}$ doping concentration in the inner layer up to 95 mol% without significant concentration quenching [26], while substantially enhancing $Tm^{3+}$ emission at the target blue range. Second, it prevents the formation of irregular and large $NaYbF_4$ nanocrystals that typically occur at high $Yb^{3+}$ doping concentration [26]. Figure **1e** presents the TEM images of the as-synthesized nanocrystals, revealing the monodisperse particles with a uniform size of ~25 nm. The high-resolution TEM image (see **Note-1** in Supplementary Materials) exhibits clear lattice fringes of {110} with a d-spacing of 0.52 nm. The Fast Fourier Transform diffraction pattern confirms the single-crystalline nature of those $Yb^{3+}/Tm^{3+}$ co-doped nanocrystals, which matches with that of hexagonal-phase NaYF$_4$ (JCPDS#16-0334), as further verified by X-ray diffraction spectrum (**Note-1** in Supplementary Materials). Then the nanocrystal suspension was directly spin-coated onto glass substrate, forming uniform thin UCNCs-film (Figure **1d**) after cyclohexane evaporation at room temperature.

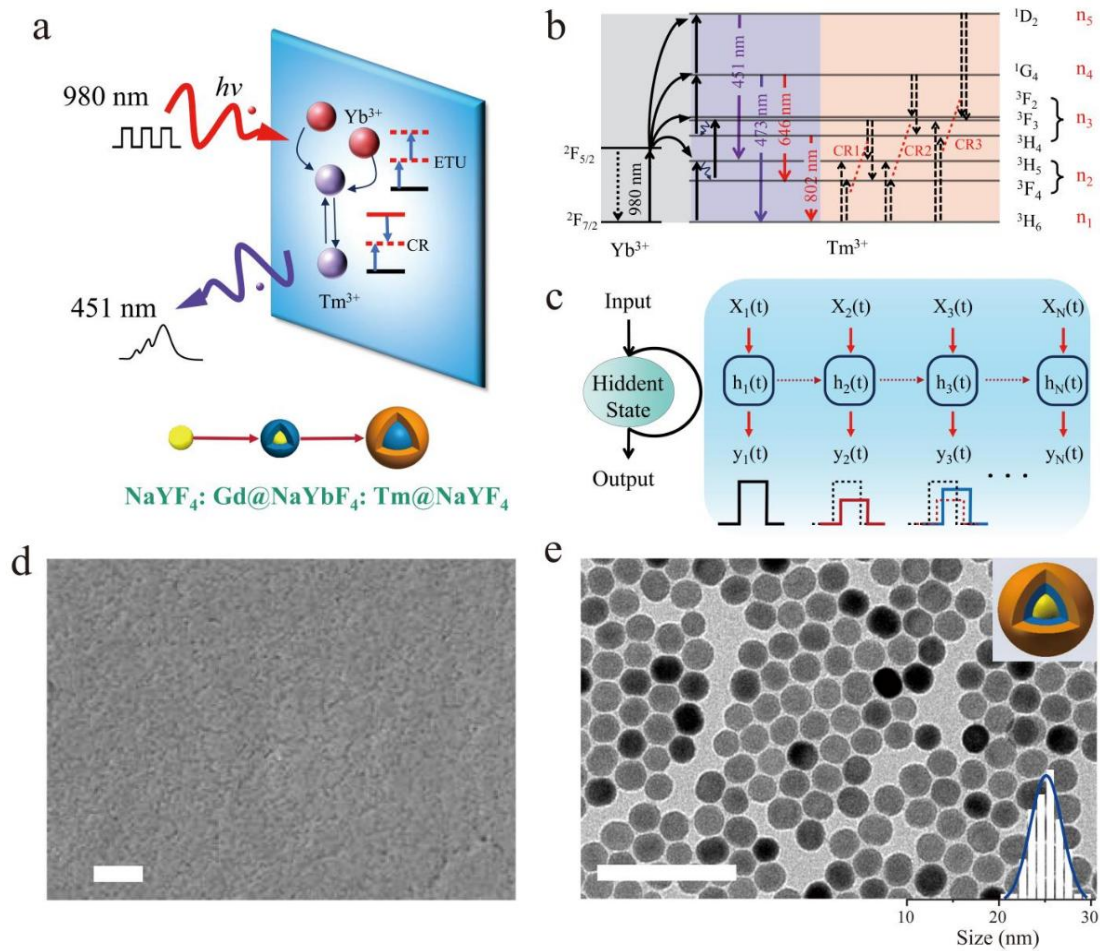

**Figure 1. The concept of UCNCs-based RNN. (a)** Schematic diagram of information processed by the UCNCs-film. **(b)** Energy level of Re$^{3+}$ ions. **(c)** Interconnections among virtual nodes within the Re$^{3+}$ nanocrystals. *X(t)* is the input data, corresponding to modulated 980 nm laser signal, *y(t)* is the output data, corresponding to the upconverted blue luminescence, and *h(t)* is the hidden states, corresponding to the luminescence dynamics of UCNCs film. **(d)** SEM image of the thin UCNCs film. (e) TEM images and particle size distribution of the core-shell-shell UCNCs.

In the NaYF$_4$:Gd@NaYbF$_4$:Tm@NaYF$_4$ sandwiched structure, Yb$^{3+}$ ions serve as sensitizers with wide absorption cross-section at 980 nm[26], while Tm$^{3+}$ ions function as activators whose luminescent energy levels can be categorized into five distinct states: n1($^3H_6$), n2($^3H_5$, $^3F_4$), n3($^3F_2$, $^3F_3$, $^3H_4$), n4($^1G_4$), n5($^1D_2$), as shown in Figure **1b**. Obviously, upon continuous-wave (CW) 980 nm excitation, efficient emission can be obtained at the center peaks of 451, 476, 646, and 802 nm, which can be assigned to $^1D_2 \rightarrow ^3H_5$, $^1G_4 \rightarrow ^3H_6$, $^1G_4 \rightarrow ^3F_4$, $^3H_4 \rightarrow ^3H_6$ transition of Tm$^{3+}$ ions (see

Figure **1b**), respectively. In the case, three groups of representative CR processes for Tm$^{3+}$ ions can be identified as CR1 (n1→n2, n3→n2), CR2 (n1→n2, n4→n3), CR3 and (n1→n3, n5→n3), respectively [27,28]. These dynamic processes can be quantitatively described by the following rate equations [18]:

$$\begin{cases} \frac{dn_{s_1}}{dt} = -P_{in}n_{s_1} + W_s n_{s_2} + (c_1 n_1 + c_2 n_2 + c_3 n_3 + c_4 n_4)n_{s_2} \\ \frac{dn_{s_2}}{dt} = P_{in}n_{s_1} - W_s n_{s_2} - (c_1 n_1 + c_2 n_2 + c_3 n_3 + c_4 n_4)n_{s_2} \end{cases} \quad (1)$$

$$\begin{cases} \frac{dn_1}{dt} = (-c_1 n_{s_2} n_1) + (b_{21} W_2 n_2 + b_{31} W_3 n_3 + b_{41} W_4 n_4 + b_{51} W_5 n_5) - (k_1 n_1 n_3 + k_2 n_1 n_4 + k_3 n_1 n_5) \\ \frac{dn_2}{dt} = (c_1 n_{s_2} n_1 - c_2 n_{s_2} n_2) + (-W_2 n_2 + b_{32} W_3 n_3 + b_{42} W_4 n_4 + b_{52} W_5 n_5) + (2k_1 n_1 n_3 + k_2 n_1 n_4) \\ \frac{dn_3}{dt} = (c_2 n_{s_2} n_2 - c_3 n_{s_2} n_3) + (-W_3 n_3 + b_{43} W_4 n_4 + b_{53} W_5 n_5) + (-k_1 n_1 n_3 + k_2 n_1 n_4 + 2k_3 n_1 n_5) \\ \frac{dn_4}{dt} = (c_3 n_{s_2} n_3 - c_4 n_{s_2} n_4) + (-W_4 n_4 + b_{54} W_5 n_5) + (-k_2 n_1 n_4) \\ \frac{dn_5}{dt} = (c_4 n_{s_2} n_4) + (-W_5 n_5) + (-k_3 n_1 n_5) \end{cases} \quad (2)$$

Equations (1) and (2) describe the luminescence processes of Yb$^{3+}$ and Tm$^{3+}$ ions, respectively. Here, $n_{s_i}$ represent the population densities at energy levels $s_i$ in Yb$^{3+}$, and $n_j$ represent the population densities at energy levels $j$ in Tm$^{3+}$. The term $P_{in}$ denotes the stimulated absorption rate of Yb$^{3+}$ ions under optical pumping, which is proportional to the pumping density. The parameter $c_i$ characterizes the energy transfer upconversion (ETU) process from $s_2$ to $n_{i+1}$, while $b_{jk}W_j$ describes the transition process from energy level $j$ to $k$. The terms $k_1, k_2, k_3$ correspond to the three possible CR processes. It should be noted that both ETU and CR processes are fundamentally non-radiative energy transfer mechanisms between neighboring ions [29], whose efficiencies are governed by several critical factors including interionic distance, spectral overlap, and doping concentration. Consequently, these processes are simultaneously influenced by multiple energy level populations, manifesting as quadratic terms in the rate equation model. This nonlinear dependence gives rise to strongly nonlinear behavior between the emission intensity of specific luminescence bands and the pumping power, thus establishing the fundamental physical basis for implementing all optical nonlinear neural networks.

To systematically characterize the optical neuron-like properties of the Re$^{3+}$ UCNCs, three key aspects must be experimentally verified [30]: (i) nonlinear power-dependent response to single optical pulses; (ii) temporal integration capability - the ability to accumulate energy from consecutive input pulses within a specific time window and to trigger nonlinear responses; and (iii) graded response characteristics - the capacity to generate incrementally enhanced outputs upon continuous stimulation. These features collectively determine whether the material exhibits essential neuromorphic functionalities. We first conducted luminescence characterization of the sample using the optical setup illustrated in Figure **2a**. We employed an external modulation system comprising a lithium niobate electro-optic modulator (EO-AM-R-20-C1, Thorlabs) and two orthogonal linear polarizers to modulate the 980 nm CW pumping laser, enabling precise control of pump intensity, pulse width, and pulse interval (**Note-2** in Supplementary Materials). Figure **2b** presents the emission spectra of the sample under various pump densities. At low pump powers (<20 mW/cm²), the $^1D_2 \rightarrow ^3H_5$ transition at 451 nm was virtually absent, with only the

$^3H_4 \rightarrow {}^3H_6$ transition at 802 nm being observable. As the excitation power increased, the 451 nm emission intensity gradually surpassed that at 802 nm, becoming the dominant upconversion luminescence. We attribute this transition to the rapid population buildup at the $^1D_2$ level through efficient ETU between $Yb^{3+}$ and $Tm^{3+}$ ions, facilitated by fast CR processes (see **Note-3** in Supplementary Materials). The inset in Figure **2b** shows the power-dependent intensity relationship for different emission wavelength. Notably, there exhibits a significant nonlinear growth along with the increasing pump power at 451 nm, while other wavelength remains essentially linear response. To enhance the optical nonlinearity for subsequent measurements, we incorporated a narrowband filter in front of the photon detector to selectively transmit photons (451 nm) from this specific transition.

Figure **2c** presents the temporal luminescence response at 451 nm under a rectangular pulse excitation (fixed pulse width: 1700 μs, pump power: 40 mW). It can be seen the luminescence intensity gradually accumulates as pulse onset until reaching saturation. And the luminescence persists after pulse termination and shows a prolonged decay tail before vanishing. This phenomenon originates from the long-lived energy states of $Re^{3+}$ ions, confirming the fading memory behavior that enables temporal signal encoding. Figure **2d** displays the output saturation intensity under fixed pulse width excitation (1700 μs) with varying pumping powers, the inset in Figure **2d** shows the time-resolved luminescence responses. It can be seen both the saturation intensity and rise time show significant dependence on the pumping-power. It is particularly noteworthy that the complex relaxation processes in $Re^{3+}$ ions create interdependent effects between pumping intensity and pulse width, which means these two non-orthogonal degrees of freedom jointly determine the final luminescence output. This interdependence is visualized in Figure **2e** through three-dimensional histograms plotting maximum luminescence intensity versus both pumping intensity (at fixed pulse width) and pulse width (at fixed pumping intensity), demonstrating nonlinear mapping relationships in both parameter spaces.

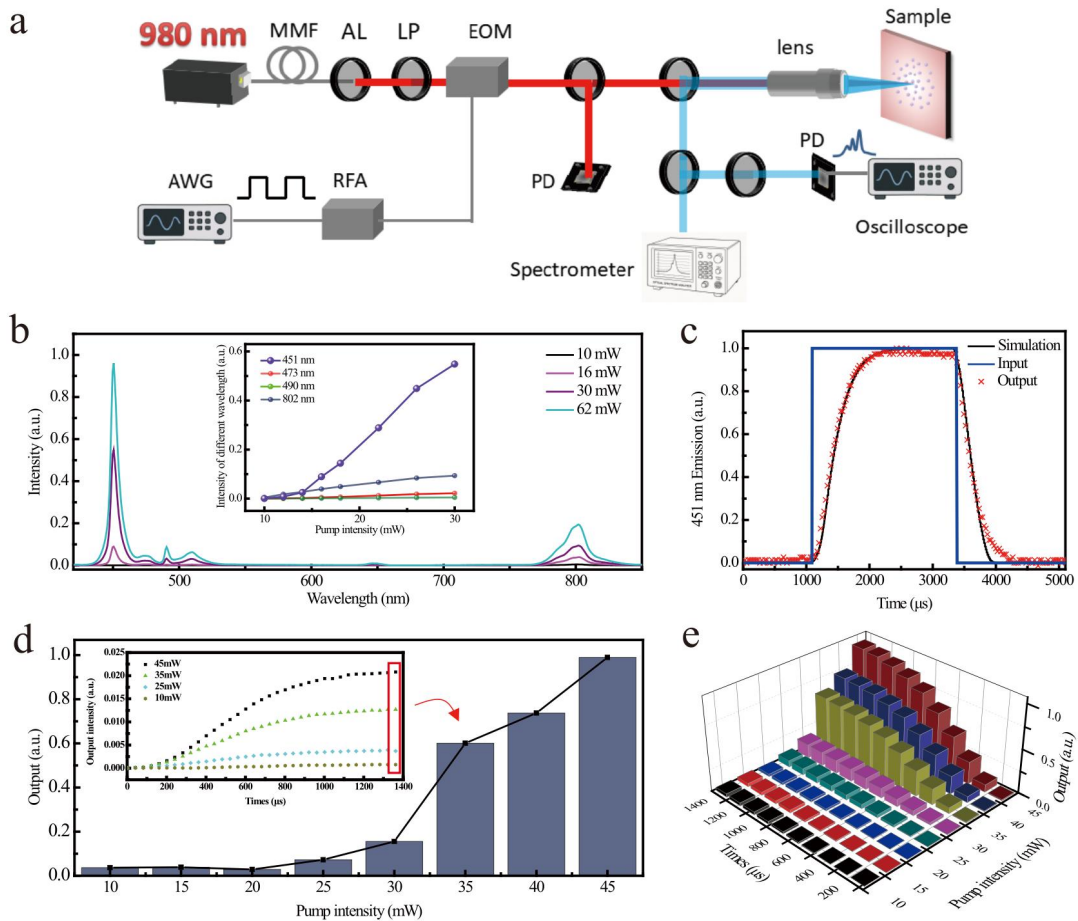

**Figure 2. Nonlinear luminescence response of UCNCs film under single pulse pumping.** (a) Schematic diagram of experimental setup. AL: aspherical lens, LP: linear polarizer. (b) Spectral response of UCNCs film under different pumping intensities, inset shows the integrated output intensity versus different pumping intensity at various wavelength. (c) Temporal response at 451 nm under a single pulse pumping. (d) Time-domain response of luminescence intensity under different pumping intensities. (e) Combined nonlinear relationship between output intensity and pumping intensity & pulse width for UCNCs film.

We further investigated the memory retention duration (or effective integration time window) of the $Re^{3+}$ UCNCs. When the inter-stimulus interval (ISI) exceeds this duration, the material loses its capability for energy integration between consecutive pulses. In our experiments, we initially employed two identical 980 nm pulses (pumping power: 45 mW, pulse width: 25 μs) as input signals. Figure **3a** shows a 451 nm output response featuring two distinct peaks, demonstrating that the $Re^{3+}$ UCNCs can produce sustained responses to sequential inputs similar to graded neurons. The performance of such neuromorphic nodes is typically characterized using paired-pulse facilitation (PPF), where the PPF index is defined as the ratio of the second peak amplitude ($A_2$) to the first ($A_1$). As shown in Figure **3b**, the PPF index decreases from 150% to 123% as Δt increases from 2.5 μs to 50 μs. The decay follows a single exponential trend with a fitted time constant of 55 μs. Notably, even at Δt = 90 μs, the PPF remains at 110%, indicating an effective integration window of approximately 90 μs.

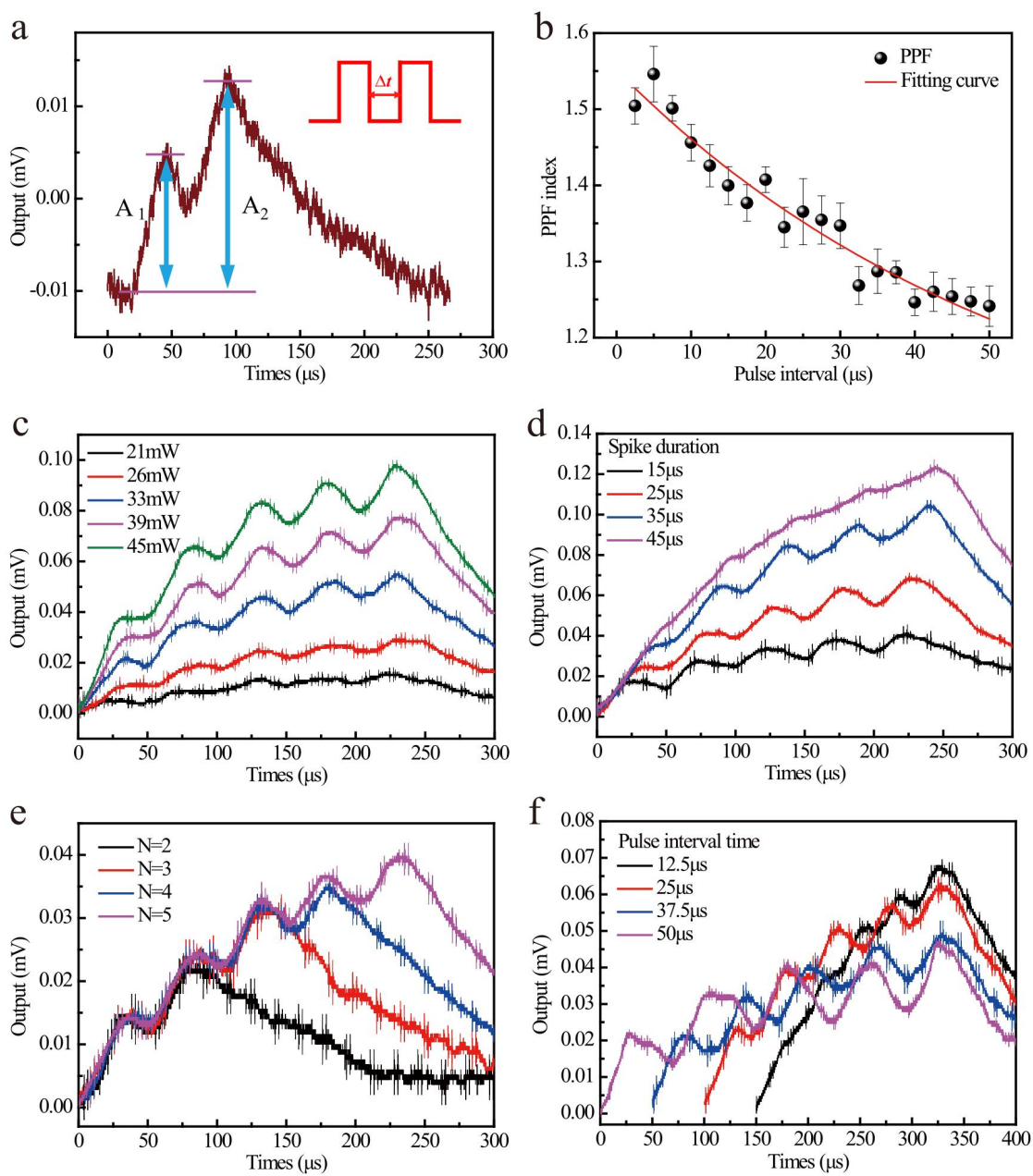

**Figure 3. Nonlinear luminescence responses of Re$^{3+}$ UCNCs film under continuous pulses pumping.** (a) PPF response of the material. (b) Variation of PPF index with inter-pulse time interval (Δt). (c) Graded luminescence response of the UCNCs film to five consecutive pulses at different pump powers with fixed pulse width (25 μs) and period (50 μs). (d) Graded luminescence response under varying pulse duration with fixed pump power (45 mW) and pulse period (50 μs). (e) Graded luminescence response of UCNCs film with fixed pump power (45 mW), pulse period (50 μs), and duration (25 μs), while adjusting the number of input pulses. (f) Graded luminescence response to varying pulse intervals at fixed pulse duration (25 μs).

We then applied five consecutive pulses (fixed width: 25 μs, period: 50 μs) and measured the 451 nm output intensity to explore the luminescence dynamics under multi-pulse stimulation. In Figure **3c**, it can be seen that despite the input signal terminating at 225 μs, the memory retention effect sustains luminescence for an extended duration (~300 μs). By increasing the pump power from 21 mW to 45 mW, we observed enhanced memory retention duration, with the output signal amplitude rising from 15.6 μV to 97.6 μV. This transition from short-term potentiation (STP) to long-term potentiation (LTP) demonstrates the power-dependent memory modulation in $Re^{3+}$ ions. Additionally, we systematically examined the influence of pulse width, pulse number, and pulse interval on the luminescence dynamics under continuous multi-pulse excitation (Figures **3d-f**). Figure **3d** displays the output intensity under fixed pump power (45 mW) and period (50 μs) while varying the pulse width from 15 μs to 45 μs. The results show that wider pulses enhance the overall emission intensity while maintaining graded memory retention. Figure **3e** demonstrates the temporal response under varying pulse numbers at fixed power, width, and interval. Increasing pulse numbers within the memory window strengthens both the retention capability and duration. Finally, figure **3f** reveals that shorter pulse intervals (ranging from 12.5 μs to 50 μs at fixed 25 μs width) yield higher peak output intensities, though with reduced memory persistence.

**Classification Tasks from $Re^{3+}$ UCNCs based Reservoir Computing**

As shown above, the luminescence intensity of $Re^{3+}$ UCNCs under multi-pulse excitation is governed by a combination of factors, including pump power, pulse width, inter-pulse interval, etc. By encoding information through these physical parameters, we constructed a $Re^{3+}$ UCNCs based reservoir computing system and applied it to classification and prediction tasks. Figure **4a** illustrates the experimental configuration, which consists of an input layer, reservoir layer, and readout layer. In the input layer, each image from the MNIST handwritten digit dataset [31] is processed as an 8-bit grayscale image. During preprocessing stage, the pixel values undergo binarization to obtain a 28*28 binary matrix of "0"s and "1"s. The reservoir layer implements a 4-bit graded neuron architecture. The input image is first segmented and rearranged using a 2*2 sliding window kernel (see **Note-4** in Supplementary Materials).

The resulting signals drive an electro-optic modulator that encodes them onto the optical power of the 980 nm pumping laser, which in turn excites the $Re^{3+}$ UCNCs. Binary "1" is represented by optical pulses of 45 mW power, 25 μs width, and a 25 μs inter-pulse interval. Then a detector measured output intensity immediately after the fourth pulse. For continuous 4-bit inputs, there be 16 possible signal encoding combinations, ranging from '0000' to '1111'. Then all 16 possible 4-bit input codes should elicit distinct nonlinear optical responses, as shown in Figure **4b**. Further analysis in Figure **4c** examines the dynamic changes in "excitatory postsynaptic current" (EPSC) under different optical pulse combinations. The ΔEPSC variations exhibit unique patterns for distinct input sequences, indicating the system's robust spatiotemporal processing capacity and adaptive response characteristics to varying input signals.

Following nonlinear transformation by the Re$^{3+}$ UCNCs, the output intensities are processed by the readout layer for classification. The readout layer consists of a linear neural network performing linear regression, with cross-entropy function employed as the loss function during training. Experimental evaluation using the standard MNIST handwritten digit datasets yielded a classification accuracy of 90.7% on 10,000 test images, as represented by the confusion matrix in Figure **4d**. Detailed performance analysis reveals an average F1-score of 0.91 (see **Note-4** in Supplementary Materials), with optimal classification achieved for digits "0", "1", and "6". Comparatively weaker performance was observed for digits "5", "8", "9", and "2", which we attribute to their structural complexity (particularly for "5" and "8" with intersecting strokes) and greater writing style variations in the MNIST datasets, leading to reduced separability in feature space.

Notably, the input-output characteristics of the Re$^{3+}$ UCNCs system exhibit persistent luminescence after the pump is turned off. These relaxation dynamics indicate that the graded neuron sustains measurable outputs throughout the decay interval, enabling continued classification within this temporal window. As shown in Figure **4e**, the classification accuracy declines gradually with increasing relaxation time-primarily due to a reduced signal-to-noise ratio from amplitude attenuation-yet remains nearly constant over the first ~40 μs. This temporal robustness suggests that practical implementations can tolerate looser timing precision in the detection stage without compromising reliability. The persistence of classification performance during relaxation underscores the system's inherent fault tolerance and operational flexibility for real-world applications.

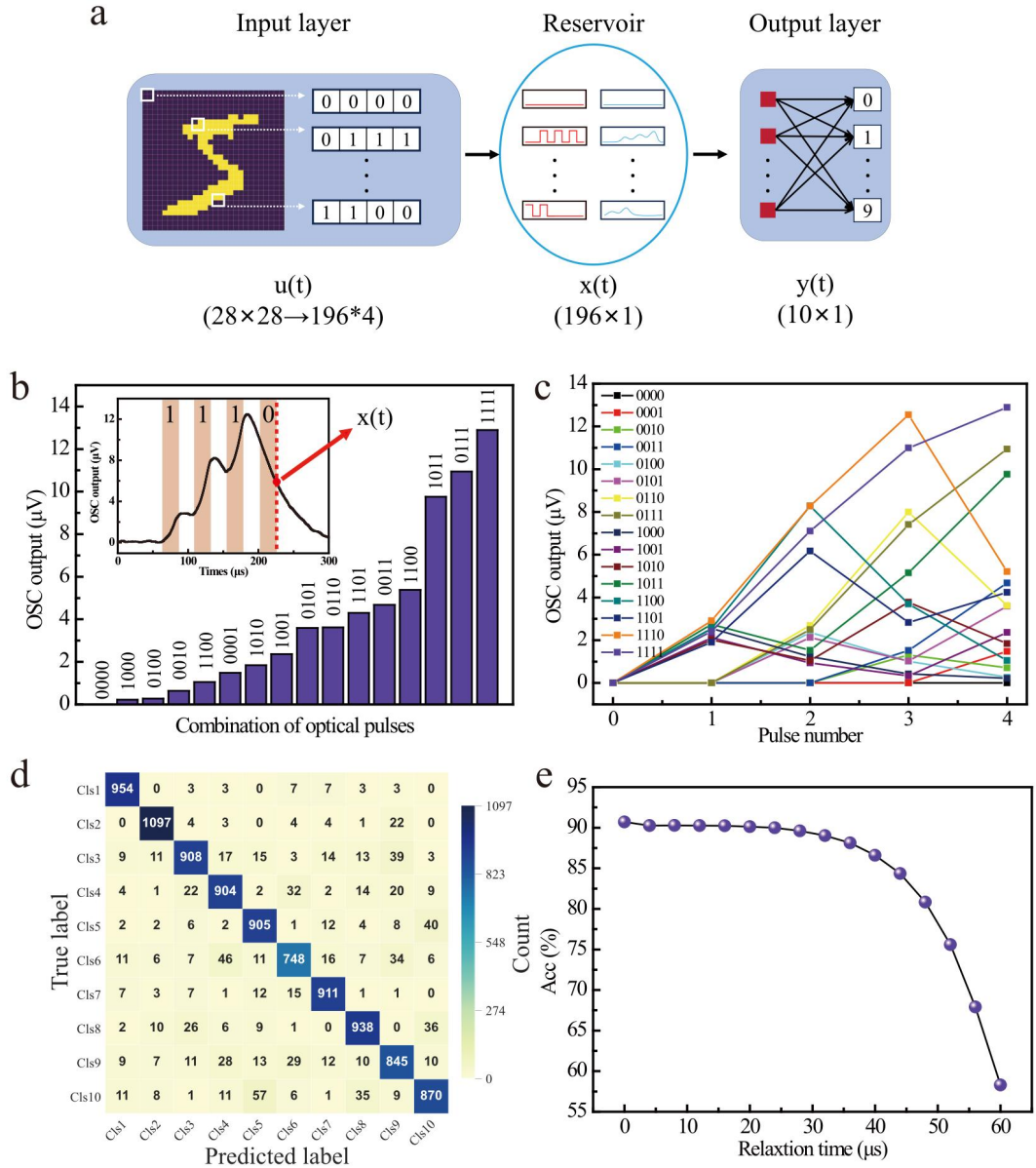

**Figure 4. Classification Tasks from Re³⁺ UCNCs based Reservoir Computing. (a)** Schematic of the graded-neuron reservoir computing architecture based on the Re$^{3+}$ UCNCs. **(b)** Distinct output intensities of the sample produced by the 4-bit binary optical pulses. **(c)** Output intensity evolution ("ΔEPSC") of the 16 different states over time. **(d)** Confusion matrix for the designed all optical RC. **(e)** Classification accuracy versus relaxation time after pumping termination.

**Chaotic Time Series Prediction Using Re³⁺ UCNCs based Reservoir Computing**

To further evaluate the temporal processing capabilities of the system, we implemented a chaotic time series prediction task based on the Mackey-Glass (MG) equation, which is the benchmark model for testing nonlinear dynamical system forecasting. Figure **5a** schematically illustrates the neural network architecture for the temporal processing task. The MG time series was generated through numerical integration of the delay differential equation:

$$\frac{dS(t)}{dt} = \alpha \frac{S(t-\tau)}{1+S(t-\tau)^\beta} - \gamma S(t) \qquad (3)$$

where *t* represents the current time step and $\tau$ denotes the time delay that characterizes the system's memory capacity for historical information. The magnitude of $\tau$ critically determines the system's chaotic behavior. The parameter $\beta$ governs the nonlinearity order, modulating the degree of nonlinear characteristics in the system dynamics. The coefficients $\alpha$ and $\gamma$ correspond to the feedback gain and decay rate, respectively. In our implementation, these parameters were systematically configured as $\alpha$=0.2, $\gamma$=0.1, $\beta$=10, and $\tau$=17.

During the data preprocessing stage, the original MG sequence is first discretized to obtain $S(t) \in R^{1\times L}$. Typically, before being input to the reservoir layer, the data is processed using a mask matrix to stimulate richer internal states of the reservoir. Here, we selected an exponential pulse mask, as shown in Figure **5a**, and set $W_{in} \in R^{N\times 1}$. Subsequently, we obtained the input signal matrix $u = W_{in} \cdot S(t) \in R^{N\times L}$, where each column represents the input signal vector of the system at time *t*. The input signals were then encoded into the intensity of 980 nm pumping light and injected into the aforementioned $Re^{3+}$ UCNCs film. The corresponding output light intensity response sequence *U(t)* could be read out by the photo-detector and oscilloscope, and subsequently used for final prediction through ridge regression. The predicted values *ŷ(t)* of the system are obtained through linear combination of the output states. The trained $W_{out}$ represents the output weights, which constitute the only parameters requiring training in the network. To quantitatively evaluate the prediction performance, we calculated the normalized root mean square error (NRMSE) using the following formula:

$$NRMSE = \sqrt{\frac{\sum_{n=0}^{L-1}(\hat{y}(n)-y(n))^2}{L\sigma^2}} \qquad (4)$$

Here, $\hat{y}(n)$ represents the predicted values, $y(n)$ denotes the target values, $L$ corresponds to the total length of the original Mackey-Glass time series, and $\sigma^2$ is the variance of the target sequence. The NRMSE metric provides a normalized measure of prediction accuracy, where smaller values indicate better predictive performance, signifying that the predicted sequence more closely approximates the ideal original time series. In addition, in reservoir computing, the reservoir state at the current time step is typically used for the readout [32]. While the network implicitly retains information from previous time steps, we can explicitly incorporate state variables from earlier time steps $\{X(t-k)\}_{k=1}^{n}$ to capture more historical information (where *n* denotes the total number of preceding time steps used). We also consider multi-step-ahead prediction, defining the forecast horizon *i* as the future – current interval between the present time *t* and the target time *t+i*.

It should be emphasized that when the Re$^{3+}$ UCNCs system functions as the reservoir layer, its output response is theoretically governed by solutions to its intrinsic rate equations of luminescence dynamics. Specifically, when a sequence of random signals is injected into the Re$^{3+}$ UCNCs, the experimentally measured output intensity series can be precisely determined through numerical simulation of the rate equations. As demonstrated in Figure **5b**, we injected a series of stochastic pulses and experimental measurements confirmed excellent agreement between the solutions of these dynamical equations and the system's actual output responses.

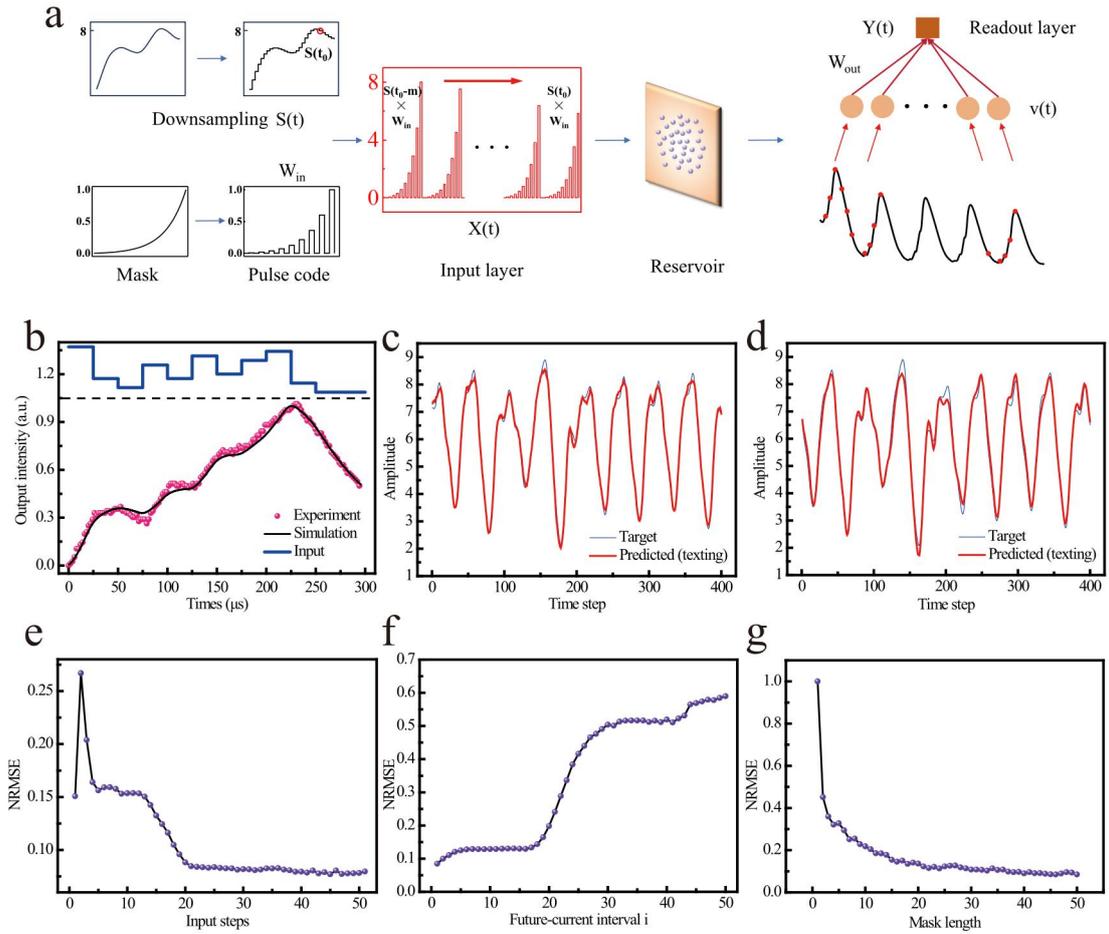

**Figure 5. Chaotic time-series prediction task based on Re$^{3+}$ UCNCs based reservoir computing**. (**a**) Schematic of the signal encoding method for the MG series and reservoir computing architecture. (**b**) Experimental measured and simulated output intensity series of Re$^{3+}$ UCNCs. (**c**), (**d**) Prediction results with future-current interval $i$ = 1 and 17, respectively. (**e**), (**f**), (**g**) The functions of NRMSE on the test set regarding multiple time steps $n$, future-current interval $i$, and mask length $N$, respectively.

To thoroughly investigate the performance of the Re$^{3+}$ UCNCs based reservoir and evaluate its temporal prediction capability under various parameter conditions, we then directly employed the numerical simulations based on the $Re^{3+}$ rate equations for all subsequent training and testing procedures. In the experiment, we generated a total of 2000 time steps for evaluation, divided into three distinct phases: the first 200 time steps served as the warm-up period (to initialize reservoir states), time steps 200-1400 constituted the training phase, and time steps 1400-2000 were reserved as the test set. As examples, figures **5c** and **5d** display the evolution curves of the predicted sequences and the ideal MG sequences for single-step predictions at $i$=1 and 17, respectively. It can be observed that the predicted curves closely approximate the ideal target sequences, remaining smooth overall without significant high-frequency noise. We then systematically investigated the impacts of the number of historical steps $n$, the future-current interval $i$, and the mask length $N$ on the system's temporal prediction performance. Figure **5e** shows the evolution curve of the NRMSE for single-step prediction tests using reservoir node states from multiple time steps $n$. At $n$=1, the NRMSE is 0.15, but as the number of reservoir node states increases, the NRMSE rapidly decreases, reaching 0.088 at $n$=20. Further increasing $n$ does not lead to significant changes in NRMSE, which stabilizes around 0.08. Figure **5f** presents the multi-step prediction results using reservoir node states from $n$=20 time steps. At intervals $i$=1 and 17, the NRMSE values are 0.084 and 0.133, respectively, indicating the system's strong multi-step prediction capability. However, beyond 17 time steps, the NRMSE rises significantly. This is because the selected Mackey-Glass sequence has an inherent feedback delay of 17 steps, leading to degraded prediction performance beyond this point. Finally, figure **5g** illustrates the impact of mask length $N$ on the reservoir network's prediction performance. It can be seen that increasing the mask length beyond $N$=20 does not significantly improve prediction accuracy but instead negatively affects computational speed.

**Discussion**

Above, we have investigated the temporal nonlinear luminescence dynamics of $Re^{3+}$ UCNCs and experimentally demonstrated a bio-inspired all-optical reservoir computing network. Compared other physical RC systems,it eliminates optoelectronic conversion losses, simplifies hardware design, offers potential for compact, parallelized reservoirs. Meanwhile, there remains room for improvement in both system processing speed and task execution accuracy (see **Note-5** in supplementary materials). In fact, the performance demonstrated in this work can be further improved in multiple aspects. At the system level, this study only demonstrates a non-delayed single-node reservoir architecture. For the physical reservoir layer, multiple feedback loops could be introduced or more complex topological network structures could be designed [33]. For example, we show that incorporating feedback loops can further enhance the temporal prediction capability of the reservoir system(**Note-4** in supplementary materials). From the device fabrication perspective, the current study only explores the luminescent properties of $Re^{3+}$ UCNCs film. Future work could employ micro/nanostructures to improve directional emission, thereby enhancing the signal-to-noise ratio of luminescence intensity [34]. Additionally, micro/nano structures could be utilized to boost emission intensity, and the Purcell effect could be exploited to shorten relaxation times of $Re^{3+}$ UCNCs, increasing system operating speed [35], among other potential improvements.

In summary, this work presents the first experimental demonstration of a bio-inspired all-optical neural network based on the intrinsic nonlinear luminescence dynamics of $Re^{3+}$ UCNCs. Benefiting from the inherent neuron-like characteristics and pronounced memory effects of these materials, The developed reservoir computing system achieves exceptional performance without requiring delayed feedback loops, excelling in both classification and time-series prediction tasks. This method eliminates the need for any additional electrical excitation or electro-optic modulators, enabling fully optical injection-based information processing while significantly reducing system complexity. Furthermore, the system circumvents the requirement for sophisticated optical feedback architectures. These features collectively open new avenues for developing next-generation low-power edge computing devices. And may applied in real-world edge scenes , such as in LiDAR, free-space optical communications, and wearable sensors[5].

## Methods

The multilayer $NaYF_4$:Gd(10 mol%)@$NaYbF_4$:Tm(5 mol%)@$NaYF_4$ nanoparticles were synthesized according to a modified method in ref[23-26]. Other experimental details are provided in the Supplementary Materials.

## Acknowledgement

The authors acknowledge support by National Key Research and Development Program of China (Grant Nos. 2024YFB2809200, 2022YFB3505700), National Natural Science Foundation of China (Grant Nos. 6233000076, 12334016, 11934012, 12025402, 12261131500, 92250302, 12474376 and 62305084), Guangdong Basic and Applied Basic Research Foundation (2023A1515011746, 2024B1515020060, 2023B1212010003, 2022A1515012108), Shenzhen Fundamental Research Projects (Grant Nos. GXWD 2022081714551), Guangdong Provincial Quantum Science Strategic Initiative(GDZX2406002, GDZX2306002), Shenzhen Science and Technology Program (Grant Nos. JCYJ20230807094401004), Shenzhen Fundamental research project (JCYJ20241202123719025, JCYJ20210324120402006, JCYJ20200109112805990, GXWD 20220817145518001, RCYX20231211090432059), the Fundamental Research Funds for the Central Universities (Grant No. HIT.OCEF.2024020, 2022FRFK01013).

## Data availability

The data that support the findings in this study is available from the corresponding authors upon reasonable request.

## Competing interests

The authors declare no competing interests.

## Author contributions

C. H. and L. J. conceived the idea and supervised the research. J.C., J.X., L.J. and K.W. prepared the experimental materials. J.C performed the experimental measurements. J.F., and J.C. did the simulation. All the authors discussed the contents and prepared the manuscript.

## Reference:

[1] Marković D, Mizrahi A, Querlioz D, et al. Physics for neuromorphic computing[J]. Nature Reviews Physics, 2020, 2(9): 499-510.
[2] Schuman C D, Kulkarni S R, Parsa M, et al. Opportunities for neuromorphic computing algorithms and applications[J]. Nature Computational Science, 2022, 2(1): 10-19.
[3] Kudithipudi D, Schuman C, Vineyard C M, et al. Neuromorphic computing at scale[J]. Nature, 2025, 637(8047): 801-812.
[4] Lukoševičius M, Jaeger H. Reservoir computing approaches to recurrent neural network training[J]. Computer science review, 2009, 3(3): 127-149.
[5] Yan M, Huang C, Bienstman P, et al. Emerging opportunities and challenges for the future of reservoir computing[J]. Nature Communications, 2024, 15(1): 2056.
[6] Cucchi M, Abreu S, Ciccone G, et al. Hands-on reservoir computing: a tutorial for practical implementation[J]. Neuromorphic Computing and Engineering, 2022, 2(3): 032002.
[7] Tanaka G, Yamane T, Héroux J B, et al. Recent advances in physical reservoir computing: A review[J]. Neural Networks, 2019, 115: 100-123.
[8] Sun J, Yang W, Zheng T, et al. Novel nondelay-based reservoir computing with a single micromechanical nonlinear resonator for high-efficiency information processing[J]. Microsystems & Nanoengineering, 2021, 7(1): 83.
[9] Zhang L, Chen Y, Mao S, et al. Lead-free halide perovskite-based optoelectronic synapse for reservoir computing[J]. Chemical Engineering Journal, 2025, 506: 160106.
[10] Shastri B J, Tait A N, Ferreira de Lima T, et al. Photonics for artificial intelligence and neuromorphic computing[J]. Nature Photonics, 2021, 15(2): 102-114.
[11] Bente I, Taheriniya S, Lenzini F, et al. The potential of multidimensional photonic computing[J]. Nature Reviews Physics, 2025: 1-12.
[12] Wu N, Sun Y, Hu J, et al. Intelligent nanophotonics: when machine learning sheds light[J]. eLight, 2025, 5(1): 5.
[13] Shen Y W, Li R Q, Liu G T, et al. Deep photonic reservoir computing recurrent network[J]. Optica, 2023, 10(12): 1745-1751.
[14] Vandoorne K, Mechet P, Van Vaerenbergh T, et al. Experimental demonstration of reservoir computing on a silicon photonics chip[J]. Nature communications, 2014, 5(1): 3541.
[15] Rafayelyan M, Dong J, Tan Y, et al. Large-scale optical reservoir computing for

spatiotemporal chaotic systems prediction[J]. Physical Review X, 2020, 10(4): 041037.

[16] Yildirim M, Dinc N U, Oguz I, et al. Nonlinear processing with linear optics[J]. Nature Photonics, 2024, 18(10): 1076-1082.

[17] Wang T, Sohoni M M, Wright L G, et al. Image sensing with multilayer nonlinear optical neural networks[J]. Nature Photonics, 2023, 17(5): 408-415.

[18] Liu Y, Lu Y, Yang X, et al. Amplified stimulated emission in upconversion nanoparticles for super-resolution nanoscopy[J]. Nature, 2017, 543(7644): 229-233.

[19] Zhang H, Zhang H. Rare earth luminescent materials[J]. Light: Science & Applications, 2022, 11(1): 260.

[20] Appeltant L, Soriano M C, Van der Sande G, et al. Information processing using a single dynamical node as complex system[J]. Nature Communications, 2011, 2(1): 468.

[21] Larger L, Soriano M C, Brunner D, et al. Photonic information processing beyond Turing: an optoelectronic implementation of reservoir computing[J]. Optics Express, 2012, 20(3): 3241-3249.

[22] Vettelschoss B, Röhm A, Soriano M C. Information processing capacity of a single-node reservoir computer: an experimental evaluation[J]. IEEE Transactions on Neural Networks and Learning Systems, 2021, 33(6): 2714-2725.

[23] Zhou, B.; Huang, J.; Yan, L.; Liu, X.; Song, N.; Tao, L.; Zhang, Q. Probing Energy Migration through Precise Control of Interfacial Energy Transfer in Nanostructure. Advanced Materials 2019, 31, e1806308.

[24] Jin, L.; Chen, X.; Wu, Y.; Ai, X.; Yang, X.; Xiao, S.; Song, Q. Dual-wavelength switchable single-mode lasing from a lanthanide-doped resonator. Nature communications 2022, 13:1727.

[25] Sun, T.; Chen, B.; Guo, Y.; Zhu, Q.; Zhao, J.; Li, Y.; Chen, X.; Wu, Y.; Gao, Y.; Jin, L.; Chu, S. T.; Wang, F. Ultralarge anti-Stokes lasing through tandem upconversion. Nature communications 2022, 13:1032.

[26] Chen, X.; Jin, L.; Kong, W.; Sun, T.; Zhang, W.; Liu, X.; Fan, J.; Yu, S. F.; Wang, F. Confining energy migration in upconversion nanoparticles towards deep ultraviolet lasing. Nature communications 2016, 7: 10304.

[27] Huang, J.; Yan, L.; An, Z.; Wei, H.; Wang, C.; Zhang, Q.; Zhou, B. Cross Relaxation Enables Spatiotemporal Color-Switchable Upconversion in a Single Sandwich Nanoparticle for Information Security. Advanced Materials 2023, 36:.

[28] Deng, R.; Qin, F.; Chen, R.; Huang, W.; Hong, M.; Liu, X. Temporal full-colour tuning through non-steady-state upconversion. Nat Nanotechnol 2015, 10, 237-242.

[29] Chen Z, Wang W, Kang S, et al. Tailorable upconversion white light emission from $Pr^{3+}$ single-doped glass ceramics via simultaneous dual-lasers excitation[J]. Advanced Optical Materials, 2018, 6(4): 1700787.

[30] Nie Y, Yang B, Wang D, et al. Integrated laser graded neuron enabling high-speed reservoir computing without a feedback loop[J]. Optica, 2024, 11(12):

1690-1699.

[31] LeCun Y, Bottou L, Bengio Y, et al. Gradient-based learning applied to document recognition[J]. Proceedings of the IEEE, 2002, 86(11): 2278-2324.

[32] Jaeger H. The "echo state" approach to analysing and training recurrent neural networks-with an erratum note[J]. Bonn, Germany: German national research center for information technology gmd technical report, 2001, 148(34): 13.

[33] Tan X S, Hou Y S, Wu Z M, et al. Parallel information processing by a reservoir computing system based on a VCSEL subject to double optical feedback and optical injection[J]. Optics Express, 2019, 27(18): 26070-26079.

[34] Zheng B, Fan J, Chen B, et al. Rare-earth doping in nanostructured inorganic materials[J]. Chemical Reviews, 2022, 122(6): 5519-5603.

[35] Chen H, Jiang Z, Hu H, et al. Sub-50-ns ultrafast upconversion luminescence of a rare-earth-doped nanoparticle[J]. Nature Photonics, 2022, 16(9): 651-657.